# Graph Neural Network Prediction of Nonlinear Optical Properties

Yomn Alkabakibi[a,✉], Congwei Xie[b,c], Artem R. Oganov[a]

[a] *Skolkovo Institute of Science and Technology, Bolshoy Boulevard 30, bld. 1, 121205, Moscow, Russian Federation*

[b] *Research Center for Crystal Materials, State Key Laboratory of Functional Materials and Devices for Special Environmental Conditions, Xinjiang Key Laboratory of Functional Crystal Materials, Xinjiang Technical Institute of Physics and Chemistry, CAS, 40-1 South Beijing Road, Urumqi 830011, China*

[c] *Center of Materials Science and Optoelectronics Engineering, University of Chinese Academy of Sciences, Beijing 100049, China*

Email: yomn.alkabakibi@skoltech.ru[✉], a.oganov@skoltech.ru

**Abstract**

Nonlinear optical (NLO) materials for generating lasers via second harmonic generation (SHG) are highly sought in today's technology. However, discovering novel materials with considerable SHG is challenging due to the time-consuming and costly nature of both experimental methods and first-principles calculations. In this study, we present a deep learning approach using the Atomistic Line Graph Neural Network (ALIGNN) to predict NLO properties. Sourcing data from the Novel Opto-Electronic Materials Discovery (NOEMD) database and using the Kurtz-Perry (KP) coefficient as the key target, we developed a robust model capable of accurately estimating nonlinear optical responses. Our results demonstrate that the model achieves 82.5% accuracy at a tolerated absolute error up to 1 pm/V and relative error not exceeding 0.5. This work highlights the potential of deep learning in accelerating the discovery and design of advanced optical materials with desired properties.

## Introduction

Nonlinear optical (NLO) materials play a crucial role in modern technologies, including laser systems, telecommunications, and quantum computing [1]. The ability to predict and engineer NLO properties of crystals is essential for advancing these applications. However, the discovery of novel NLO materials remains a significant challenge in materials science. Traditional methods for identifying such materials rely heavily on trial and error, an approach that is both labor-intensive and costly when conducted experimentally. While computational methods have been developed to calculate the second harmonic response of potential materials [2, 3], these approaches often require substantial computational resources and are time-consuming.

The application of data science methods in materials science has seen remarkable progress in recent years. Modern deep learning (DL) models, such as CrabNet [4], Roost [5], M3GNet [6], and iCGCNN [7], have demonstrated exceptional accuracy in predicting material properties, including band gaps [5], energies of formation [4, 6], thermal conductivity [4], and energy [4, 7]. These advancements highlight the potential of DL to revolutionize materials discovery by enabling rapid and accurate property predictions.

Despite these successes, predicting NLO properties remains particularly challenging due to the inherent complexity of NLO phenomena. Recent advances in machine learning (ML) and deep learning (DL) offer promising avenues to address these challenges by providing efficient and accurate tools for property prediction.

In this study, we leverage the Atomistic Line Graph Neural Network (ALIGNN) [8], a powerful and widely used neural network architecture in materials science, to predict NLO properties. ALIGNN is particularly well-suited for this task due to its ability to capture complex relationships between atomic structures and material properties, including three-body interactions. We utilize the NOEMD dataset [9], which comprises over 2,200 distinct non-centrosymmetric (NCS) nonlinear crystals. Each crystal is represented in the form of a JSON-encoded pymatgen structure [10], along with its second harmonic generation (SHG) tensors, which consist of 18 components calculated using a first-principles approach [1]. The NOEMD database also includes additional descriptors, such as chemical formula, space group, number of atoms, volume, electronic bandgap (calculated using HSE06 and PBE functionals), and birefringence, providing a comprehensive foundation for training and validating our model.

In this work, we extracted the Kurtz-Perry (KP) coefficient [11] from the SHG tensor to use as a key descriptor for training our model. The Kurtz-Perry method [12], originally developed as an experimental technique, evaluates and classifies crystals in powder form based on their symmetry and nonlinear optical response. It categorizes non-centrosymmetric crystals into four classes: (1) phase-matchable for SHG with large NLO response, (2) phase-matchable for SHG with small NLO response, (3) non-phase-matchable for SHG with large NLO response, and (4) non-phase-matchable for SHG with small NLO response. Beyond its experimental utility, the Kurtz-Perry method extends to provide a general expression (1) for calculating the KP coefficient for different point groups exhibiting SHG behavior. This coefficient encapsulates the essence of the SHG tensor into a single scalar value, making it an ideal descriptor for machine learning models.

$$d_{KP}^2 = \frac{5}{7}(d_{lmn})^2 + \frac{19}{105}\sum_i (d_{iii})^2 + \frac{13}{105}\sum_{i \neq j}(d_{iii}d_{ijj}) + \frac{44}{105}\sum_{i \neq j}(d_{iij})^2 + \frac{13}{105}\sum_{ijk,\,cyclic}(d_{iij}d_{jkk}). \quad (1)$$

where the subscript *lmn* can be any cyclic combination of the indices, i.e., 123 and 312.

Using the KP coefficient and performing advanced hyperparameter tuning and cross-validation, we aim to develop a robust and generalizable model for predicting NLO properties. The primary contributions of this work are threefold: (1) the application of multi hyperparameter-tuned versions of ALIGNN to predict the KP coefficient, a key feature for determining the nonlinearity of optical crystals; (2) the integration of the NOEMD dataset and KP coefficient to enhance model performance; and (3) a comprehensive evaluation of the model's accuracy and generalizability. Our findings demonstrate the potential of deep learning to revolutionize the discovery and design of NLO materials, paving the way for accelerated materials innovation.

## Methods and Model Development

### Dataset Preprocessing

The NOEMD dataset, comprising 2,274 non-centrosymmetric (NCS) nonlinear crystals, was split into training, validation, and test sets in the ratio 8:1:1. This partitioning ensured a robust evaluation of the model's performance while maintaining a sufficiently large training set for effective learning.

To address the wide range of the KP coefficient (0.05–300 pm/V), which spans several orders of magnitude, we applied a logarithmic scale, yielding a more manageable distribution of the target variable, $ln(d_{kp})$. The log-transformed KP coefficient was used as the target for training the model, enabling smoother convergence during the learning process. However, the final model's performance evaluation was conducted on the basis of KP actual values.

**Model and Hyperparameter Tuning**

We employed the Atomistic Line Graph Neural Network (ALIGNN), a state-of-the-art graph neural network architecture designed for materials science applications. As per design, ALIGNN takes as input the crystal structure, and outputs the predicted $ln(d_{kp})$ value.

To optimize the performance of the ALIGNN model, we conducted a systematic hyperparameter tuning process using a fixed split of the dataset into training, validation, and test sets (8:1:1 ratio). We employed a manual grid search approach, where all but one hyperparameter were fixed, and the remaining one was varied to assess its impact on model performance. This process was repeated sequentially for each hyperparameter to identify the optimal combination. Three key hyperparameters were investigated: (1) the number of ALIGNN layers, (2) the learning rate (LR), and (3) the weight decay (WD). The goal was to identify the combination of hyperparameters that minimized validation loss while avoiding overfitting and ensuring stable training dynamics.

- **Number of ALIGNN Layers**: We evaluated configurations with varying numbers of layers, including: $2, 3, 4\ (default), 5\ and\ 6$ layers. This allowed us to assess the impact of model depth on performance and generalization.
- **Learning Rate (LR)**: We tested learning rates across a range of values, including: $1 \times 10^{-6}, 1 \times 10^{-5}, 1 \times 10^{-4}, 1 \times 10^{-3}\ (default), and\ 1 \times 10^{-2}$. Lower learning rates were prioritized to ensure stable gradient descent and avoid large fluctuations in the loss function.
- **Weight Decay (WD)**: Weight decay was tuned to control regularization and prevent overfitting. Tested values included: $0, 1 \times 10^{-7}, 1 \times 10^{-6}\ (default), 1 \times 10^{-5}, 1 \times 10^{-4}, and\ 1 \times 10^{-3}$. This range allowed us to balance model complexity and generalization performance.

After extensive experimentation using this manual grid search approach on the fixed dataset split, we found that a combination of a low learning rate ($1 \times 10^{-5}$), low weight decay ($1 \times 10^{-5}$), and the default 4 ALIGNN layers provided the most stable training process. This configuration minimized fluctuations in the loss function and enabled early detection of overfitting, as evidenced by the convergence of training and validation losses.

**Training Process**

The model was trained using the default AdamW optimizer implemented in ALIGNN, with the hyperparameters described in the previous section. Training was conducted for up to 300 epochs, with early stopping implemented to halt the process if the validation loss did not improve for a predefined patience period of 20 epochs. This approach effectively mitigated the risk of overfitting. The training process was closely monitored through the analysis of training and validation loss curves, which confirmed steady learning dynamics.

**Cross-Validation**

To evaluate the robustness and generalizability of the ALIGNN model, we employed a shuffle split cross-validation strategy. This approach involved randomly splitting the dataset into training, validation, and test sets multiple times, ensuring that each split maintained a representative distribution of the target variable, $ln(d_{KP})$. Specifically, we performed 30 iterations, with 80% of the data used for training and 10% each for validation and testing in each iteration.

A key feature of our approach was the use of averaging predictions for overlapping points across iterations. Since the model was trained from scratch in each iteration, this technique allowed us to assess the stability of predictions for data points that appeared in multiple test sets. Importantly, this averaging process did not impact model performance, as each iteration involved independent training. This method was chosen to validate whether a committee voting technique (i.e., aggregating predictions across several trained models) could improve model performance in this context.

The cross-validation results demonstrated that the ALIGNN model achieved consistent performance across all iterations, with a Mean Absolute Error (MAE) of 0.56 pm/V and a standard deviation of 0.05 pm/V. Additionally, the averaging of predictions for overlapping points across multiple models revealed a slight improvement in overall prediction accuracy, lowering the overall MAE from 2.99 pm/V to 2.82 pm/V.

**Results and Discussion**

**Performance Evaluating**

Traditional metrics such as RMSE, MAE, and MAPE provide valuable insights into a model's overall performance but offer limited granularity when evaluating predictions for individual structures. Fig. 1, shows the overall comparison of prediction vs. true of all cross-validation iterations in logarithmic scale, with performance metrics.

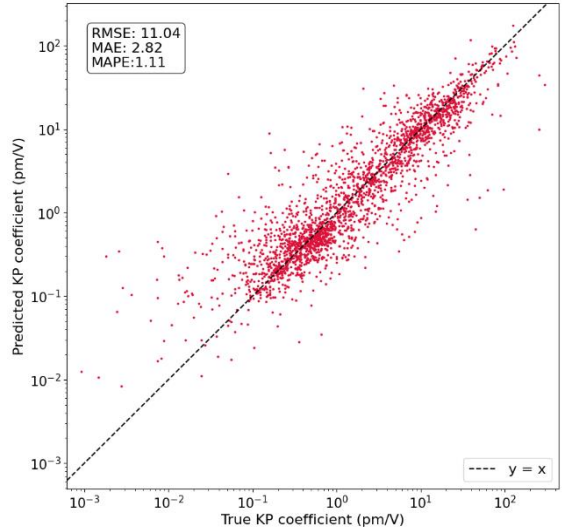

Fig. 1. Prediction vs. true of all cross-validation iterations in logarithmic scale, and performance metrics (RMSE, MAE, MAPE).

However, the limited power of traditional evaluation metrics is particularly relevant for tasks like predicting NLO properties of crystals, where the KP coefficient spans a wide range of values. To address this challenge, we developed a criterion that balances numerical accuracy and practical utility, taking into account both the absolute error and relative error of each prediction.

The absolute error $\delta_{abs}$ and relative error $\delta_{relative}$ are defined as:

$$\delta_{abs} = \left| d_{KP}^{True} - d_{KP}^{Pred} \right|, \qquad (2)$$

$$\delta_{relative} = \left| \frac{\left| d_{KP}^{True} - d_{KP}^{Pred} \right|}{d_{KP}^{True}} \right|, \qquad (3)$$

A high-confidence (HC) prediction is defined as satisfying the following logical disjunction:

$$\delta_{abs} < 1\, pm/V \quad \vee \quad \delta_{relative} < 0.5$$

Conversely, a low-confidence (LC) prediction is identified when both of the following conditions are met (logical conjunction):

$$\delta_{abs} \geq 1\, pm/V \quad \wedge \quad \delta_{relative} \geq 0.5$$

Fig. 2, shows a scatter plot of predicted versus true KP values, highlighting the distribution of HC and LC predictions. The plot demonstrates that the majority of predictions fall within the high-confidence region, with only a fraction of outliers exhibiting both intolerable relative errors and large absolute errors. LC predictions sum to 17.5% of the overall structures count, meaning that in 82.5% of structures, our model gives reliable predictions of the KP coefficient, according to the evaluation criteria discussed earlier.

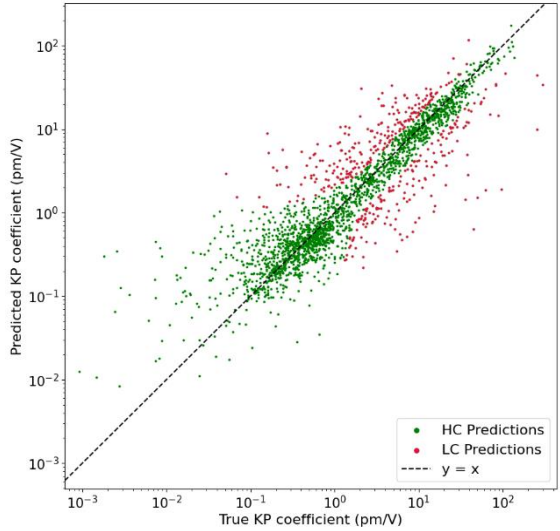

Fig. 2. Prediction vs. true of all cross-validation iterations in logarithmic scale, highlighting high confidence (HC) predictions in green, and low confidence (LC) predictions in red.

## Performance Analysis

To evaluate the model's ability to distinguish between materials with the same chemical composition but different crystal structures and symmetry, we analyzed two groups of materials: $BeAl_2Se_4$ and $LiMg_2GaSe_4$.

It is important to note that the NOEMD database includes many structures generated through a materials design approach, primarily using SMACT (Semiconducting Materials from Analogy and Chemical Theory) [13] to generate chemical compositions. As a result, the dataset used for training contains numerous materials with identical chemical compositions but differing in structure. This characteristic of some of the structures in the dataset makes it ideal to investigate the model's performance in case where differences in symmetry and structure can significantly impact the NLO properties.

Table 1: Polymorphs of $BeAl_2Se_4$ and $LiMg_2GaSe_4$ compositions, with their noemd-id, space groups, volume, $d_{KP}^{True}$, $d_{KP}^{Pred}$, $\delta_{abs}$, $\delta_{relative}$ and prediction quality

| noemd-id | Composition | Space group | Cell volume $A°$ | $d_{KP}^{True}$ $pm/V$ | $d_{KP}^{Pred}$ $pm/V$ | $\delta_{abs}$ $pm/V$ | $\delta_{relative}$ | Prediction Quality |
|---|---|---|---|---|---|---|---|---|
| noemd-4439 | $BeAl_2Se_4$ | $Cm$ | 320.56 | 7.39 | 6.96 | 0.43 | 0.06 | HC |
| noemd-4452 | | $Cm$ | 429.69 | 2.72 | 2.90 | 0.18 | 0.06 | HC |
| noemd-4442 | | $I\bar{4}2m$ | 411.35 | 2.54 | 2.22 | 0.32 | 0.13 | HC |
| noemd-4446 | | $I\bar{4}2m$ | 321.29 | 9.53 | 9.63 | 0.10 | 0.01 | HC |
| noemd-4445 | | $C2$ | 414.23 | 2.55 | 2.49 | 0.06 | 0.02 | HC |
| noemd-4448 | | $Amm2$ | 318.44 | 11.39 | 9.17 | 2.24 | 0.20 | HC |
| noemd-4453 | | $I\bar{4}$ | 319.89 | 9.13 | 9.60 | 0.47 | 0.05 | HC |
| noemd-3088 | $LiMg_2GaSe_4$ | $Pm$ | 213.89 | 8.16 | 9.40 | 1.24 | 0.15 | HC |
| noemd-3091 | | $Pm$ | 223.18 | 5.83 | 5.73 | 0.10 | 0.018 | HC |
| noemd-3092 | | $Pm$ | 204.27 | 12.68 | 12.39 | 0.29 | 0.02 | HC |
| noemd-3095 | | $Pmm2$ | 206.50 | 18.56 | 12.82 | 5.74 | 0.31 | HC |
| noemd-3102 | | $P\bar{4}2m$ | 203.35 | 8.39 | 8.66 | 0.27 | 0.03 | HC |
| noemd-3103 | | $P\bar{4}2m$ | 241.98 | 2.05 | 2.10 | 0.05 | 0.02 | HC |

As shown in Table 1, the model achieves minimal error values across all cases, demonstrating its strong predictive power. However, it is important to highlight a key trend: structures with larger NLO responses tend to exhibit higher absolute errors, while their relative errors remain within acceptable limits. This behavior is exemplified by materials such as noemd-4448 and noemd-3095, where the absolute errors are larger due to the magnitude of the true $d_{KP}$ values, but the relative errors remain relatively small.

In Table 2, we highlight the model's performance when predicting structures with exceptionally large nonlinear optical (NLO) responses. While these predictions often exhibit larger absolute errors, this behavior is primarily attributed to the underrepresentation of such materials in the dataset and overrepresentation of materials that are located on the lower end of the KP coefficient values. This natural distribution of the dataset limits the model's ability to precisely predict their KP values. Despite this limitation, the model remains highly effective in identifying materials with high KP values, which is often more critical for practical applications than achieving precise numerical accuracy.

Table 2: known phases with high NLO response, with their noemd-id, mp-id, composition, space groups, $d_{KP}^{True}$, $d_{KP}^{Pred}$, $\delta_{abs}$, $\delta_{relative}$ and prediction quality

| noemd-id | mp-id | Composition | Space group | $d_{KP}^{True}$ pm/V | $d_{KP}^{Pred}$ pm/V | $\delta_{abs}$ pm/V | $\delta_{relative}$ | Prediction Quality |
|---|---|---|---|---|---|---|---|---|
| noemd-6750 | mp-19765 | $In_2HgTe_4$ | $I\bar{4}$ | 128.12 | 103.36 | 24.76 | 0.19 | HC |
| noemd-6613 | mp-3668 | $CdGeP_2$ | $I\bar{4}2d$ | 117.80 | 95.74 | 22.06 | 0.19 | HC |
| noemd-7134 | mp-4524 | $ZnGeP_2$ | $I\bar{4}2d$ | 68.79 | 75.77 | 6.98 | 0.10 | HC |
| noemd-6462 | mp-1215555 | $Zn_2Si_2As_3P$ | $C2$ | 64.52 | 63.91 | 0.61 | 0.0095 | HC |
| noemd-6344 | mp-1224210 | $InCuGeSe_4$ | $I\bar{4}$ | 52.51 | 53.34 | 0.83 | 0.016 | HC |
| noemd-6104 | mp-12779 | $CdTe$ | $P6_3mc$ | 41.51 | 59.35 | 17.84 | 0.43 | HC |
| noemd-6549 | mp-27809 | $Ba(GeP)_2$ | $P4_2mc$ | 36.82 | 60.97 | 24.15 | 0.66 | LC |
| noemd-5694 | mp-1190 | $ZnSe$ | $F\bar{4}3m$ | 24.31 | 27.48 | 3.17 | 0.13 | HC |

## Conclusion

In this study, we demonstrated the effectiveness of the Atomistic Line Graph Neural Network (ALIGNN) in predicting the nonlinear optical (NLO) properties of crystals, particularly the KP coefficient. By leveraging the NOEMD dataset and incorporating the KP coefficient as a key descriptor, we developed a robust model capable of accurately estimating NLO responses across a wide range of materials. The model achieved a mean MAE of 0.56 pm/V with a standard deviation of 0.049 pm/V during cross-validation, highlighting its consistency and generalizability.

However, our study has some limitations. While the dataset is comprehensive, it is naturally biased toward materials with smaller nonlinear optical responses, which impacts the overall performance of our model. This bias likely affects not only our approach but also any other deep learning model applied to the same task. Addressing this imbalance in future datasets could significantly enhance the performance and generalizability of such models.

A promising direction for further research is the integration of our trained models with structure prediction algorithms, for fast identification of novel promising NLO materials.

## Data Availability

The data that support the findings of this study are available from the corresponding author upon reasonable request.

## Author Contributions

Conceptualization: Y.A., A.O.; Methodology: Y.A., A.O.; Formal Analysis: Y.A.; Writing – Original Draft: Y.A.; Writing – Review & Editing: A.O., C.X.; Supervision: A.O.

## Conflict of Interests

The authors of this work declare that they have no conflicts of interest

## Acknowledgment

The work was funded by Russian Science Foundation grant № 24-43-00162.

## Declaration of generative AI and AI-assisted technologies in the writing process

During the preparation of this work the authors used DeepSeek-R1 and DeepSeek-V3 models in order to improve the text and readability. After using this tool/service, the authors reviewed and edited the content as needed and take full responsibility for the content of the publication.